# Density and excess molar enthalpy of (2-propanol + glyme) liquid mixtures. Application of the Flory model

João Victor Alves-Laurentino[a], Fatemeh Pazoki[a], Luis Felipe Sanz[a], Juan Antonio González[a], Fernando Hevia[a], Daniel Lozano-Martín*,[a]

[a] GETEF. Departamento de Física Aplicada. Facultad de Ciencias. Universidad de Valladolid. Paseo de Belén, 7, 47011 Valladolid, Spain.

* Corresponding author, e-mail: daniel.lozano@uva.es

## Abstract

For glymes of general formula $CH_3O(CH_2CH_2O)_uCH_3$, with $u$ = 1, 2, 3, 4, the densities of the (2-propanol + glyme) systems at temperatures ranging from (293.15 to 303.15) K and at pressure 0.1 MPa were determined using a DSA 5000 densimeter (from Anton Paar). The corresponding excess molar volumes were calculated from these density measurements. In addition, excess molar enthalpies at 298.15 K and 0.1 MPa were measured using a Tian-Calvet micro-calorimeter. The results show that alkanol–ether interactions are strong but do not contribute significantly to the excess molar enthalpy, as the values are large and positive, and comparable to those of (glyme + $n$-heptane) systems. The excess molar volumes are small or even negative (in the case of the mixture with $u$ = 4), indicating that they are mainly governed by structural effects. Mixtures with 1-propanol or 2-propanol behave similarly, although interactions between unlike molecules become slightly stronger when 1-propanol is involved. On the other hand, effects related to alcohol self-association play a decisive role in the thermodynamic properties when glymes are replaced by di-$n$-propyl ether. This is supported by the application of the Flory model, which shows that orientational effects are weak in the studied glyme-containing mixtures but become significantly stronger when di-$n$-propyl ether is considered.

# 1. Introduction

The study of (alkanol + ether) mixtures has attracted significant interest due to their wide range of applications. These systems are commonly used as gasoline additives, increasing the octane rating and reducing emissions [1,2]. Solutions comprising an alkanol as a refrigerant and a polyether as an absorbent have been proposed as effective working fluids in absorption heat pumps [3]. (Alkanol + ether) systems are also relevant in other industrial processes, particularly because alkanols serve as fundamental components in the synthesis of oxaalkanes and are therefore often present as impurities in such processes. Moreover, mixtures of short-chain alkanols with linear polyethers are useful as simplified models for the complex (water + polyethylene glycols) systems, which are extensively used in biochemical and biomedical applications [4]. These solutions, especially those involving cyclic ethers, are also gaining attention in biotechnology [5,6].

From a theoretical perspective, the investigation of (alkanol + ether) mixtures is crucial due to the complex interactional and structural effects involved. In this context, when studying such systems, one must consider the self-association of alkanols and the dipolar interactions between ether molecules, both of which are disrupted upon mixing, as well as the formation of new alkanol-ether interactions during the mixing process. The presence of multiple oxygen atoms within the same ether molecule introduces an additional problem due to the so-called proximity effects, which arise from dipolar interactions between these atoms [7–9]. Consequently, the analysis of this class of solutions serves as a rigorous test for any theoretical model [9]. Furthermore, it also provides a useful preliminary step toward studying mixtures containing hydroxyethers, where intramolecular effects between the –OH and –O– groups within the same molecule become significant [10].

Our research group has extensive experience in the investigation of systems containing –OH and/or –O– groups, either within the same molecule or in different ones. In particular, we have focused on (1-alkanol + linear monoether [8,11], or + polyether [12–14] or + cyclic ether [7,11]) mixtures or to systems involving hydroxyethers [10]. As a continuation of these works, and as a contribution to the Thermodynamics of Organic Mixtures (TOM) project [15–17], we now report density measurements, $\rho$, of the (2-propanol + linear polyether) mixtures at temperatures $T$ = (293.15 to 303.15) K and pressure $p$ = 0.1 MPa. The linear polyethers considered follow the general formula $CH_3O(CH_2CH_2O)_uCH_3$, with $u$ = 1, 2, 3, 4, referred to as mono, di, tri, and tetra-glyme, respectively. The results were used to calculate the corresponding excess molar volumes, $V_m^E$. In addition, calorimetric measurements of the excess molar enthalpy, $H_m^E$, were performed at $T$ = 298.15 K and $p$ = 0.1 MPa for the same systems. This set of data is essential to achieve the objective of this work: to provide insight into the interactional and structural changes occurring in these mixtures, and to analyze the effect on the mixture properties of replacing a primary alkanol with a secondary one, namely, 1-propanol with 2-propanol, while maintaining the same number of carbon atoms.

# 2. Materials and methods

## 2.1. Materials

The liquids used in this study are listed in Table 1, along with their source and purity.

**Table 1**. Description of the compounds used in this work.

| Chemical name | CAS Number | Source | [a] Initial purity | Purification method |
|---|---|---|---|---|
| 2-propanol | 67-63-0 | Sigma-Aldrich | 0.9995 | none |
| 2,5-dioxahexane (monoglyme) | 110-71-4 | Sigma-Aldrich | 0.999 | none |
| 2,5,8-trioxanonane (diglyme) | 111-96-6 | Fluka | 0.998 | none |
| 2,5,8,11-tetraoxadodecane (triglyme) | 112-49-2 | Sigma-Aldrich | 0.999 | none |
| 2,5,8,11,14-pentaoxapentadecane (tetraglyme) | 143-24-8 | Sigma-Aldrich | 0.999 | none |

[a] Gas chromatography area fraction. Certified by the supplier.

Table 2 shows the measured properties of the pure compounds at a pressure of $p$ = 0.1 MPa: density, $\rho$, at temperatures $T$ = (293.15 to 303.15) K, and the isobaric thermal expansion coefficient, $\alpha_p$, at $T$ = 298.15 K. The values $\alpha_p$ have been obtained using the following equation:

$$\alpha_p = -\frac{1}{\rho}\left(\frac{\partial \rho}{\partial T}\right)_p \tag{1}$$

where $\rho$ is the density at $T$ = 298.15 K, and $(\partial \rho/\partial T)_p$ is calculated as the slope of the linear regression of $\rho$ as a function of $T$ in the range $T$ = (293.15 to 303.15) K. The comparison between the measured and literature values is satisfactory. For subsequent calculations, the literature values of the isothermal compressibilities, $\kappa_T$, of the pure compounds at $T$ = 298.15 K are also included in Table 2.

**Table 2.** Density, $\rho$, isobaric thermal expansion coefficient, $\alpha_p$, and isothermal compressibility, $\kappa_T$, of the pure liquids used in this work at temperature $T$ and pressure $p$ = 0.1 MPa. Root-mean-square deviation, $\sigma(F)$, of the literature values from the experimental values at temperature $T$ = 298.15 K (see equation (6)).

| Compound | [a] Property $F$ | This work | | | Literature | |
|---|---|---|---|---|---|---|
| | | $T$ = 293.15 K | $T$ = 298.15 K | $T$ = 303.15 K | $T$ = 298.15 K | [a] $\sigma(F)$ |
| 2-propanol | $\rho$/(g cm$^{-3}$) | 0.78505 | 0.78085 | 0.77660 | 0.7809 [18]<br>0.7809 [19]<br>[b] 0.78087 [20]<br>0.7809 [21]<br>0.78087 [22] | 0.000045 |
| | $\alpha_p$/(10$^{-3}$ K$^{-1}$) | | 1.082 | | 1.09 [18]<br>1.084 [19]<br>[b] 1.085 [20]<br>1.075 [21]<br>[b] 1.086 [22] | 0.0060 |
| | $\kappa_T$/(TPa)$^{-1}$ | | | | 1110 [21] | |
| monoglyme | $\rho$/(g cm$^{-3}$) | 0.86717 | 0.86170 | 0.85621 | 0.86183 [23]<br>0.86114 [24]<br>0.86114 [25]<br>0.86110 [26]<br>0.8616 [27]<br>0.86208 [28] | 0.00048 |
| | $\alpha_p$/(10$^{-3}$ K$^{-1}$) | | 1.272 | | 1.269 [23]<br>[b] 1.277 [24]<br>[b] 1.274 [25]<br>[b] 1.29 [26] | 0.011 |
| | $\kappa_T$/(TPa)$^{-1}$ | | | | 1114.5 [27] | |

| Compound | Property | | | | Literature values | U |
|---|---|---|---|---|---|---|
| diglyme | $\rho$/(g cm$^{-3}$) | 0.94370 | 0.93874 | 0.93376 | 0.93872 [23]<br>0.93935 [24]<br>0.93883 [25]<br>0.93859 [26]<br>0.9384 [29]<br>0.93882 [30] | 0.00032 |
|  | $\alpha_p$/(10$^{-3}$ K$^{-1}$) |  | 1.059 |  | 1.063 [23]<br>1.064 [24]<br>1.057 [25]<br>1.055 [26] | 0.0045 |
|  | $\kappa_T$/(TPa)$^{-1}$ |  |  |  | 821.6 [27] |  |
| triglyme | $\rho$/(g cm$^{-3}$) | 0.98485 | 0.98010 | 0.97535 | 0.98079 [23]<br>0.98058 [24]<br>0.98049 [31]<br>0.9805 [32]<br>0.98064 [33]<br>0.98000 [34] | 0.00051 |
|  | $\alpha_p$/(10$^{-3}$ K$^{-1}$) |  | 0.969 |  | 0.969 [23]<br>0.969 [24]<br>b 0.969 [31]<br>b 0.964 [32]<br>b 0.969 [33] | 0.0025 |
|  | $\kappa_T$/(TPa)$^{-1}$ |  |  |  | 707.1 [27] |  |
| tetraglyme | $\rho$/(g cm$^{-3}$) | 1.01090 | 1.00626 | 1.00164 | 1.00569 [23]<br>1.00668 [24]<br>1.00716 [35]<br>1.00568 [36]<br>1.0066 [34] | 0.00066 |
|  | $\alpha_p$/(10$^{-3}$ K$^{-1}$) |  | 0.920 |  | 0.925 [23]<br>0.928 [24]<br>b 0.915 [35]<br>b 0.929 [36] | 0.0081 |
|  | $\kappa_T$/(TPa)$^{-1}$ |  |  |  | 644.0 [37] |  |

a Expanded uncertainties ($U_1$), with a coverage factor of 2, evaluated under repeatability conditions of measurement: $U_1(T) = 0.02$ K, $U_1(p) = 10$ kPa, and $U_1(\rho) = 0.00010$ g cm$^{-3}$, $U_1(\alpha_p) = 0.020 \cdot 10^{-3}$ K$^{-1}$. Expanded uncertainties ($U_2$), with a coverage factor of 2, evaluated under reproducibility conditions of measurement: for $F = \rho, \alpha_p$, $U_2(F) = 2\sigma(F)$. Total expanded uncertainty ($U$): $U(F) = [U_1(F)^2 + U_2(F)^2]^{1/2}$.

b Calculated from data reported in the reference.

## 2.2. Apparatus and procedure

Density data were determined using a vibrating-tube densimeter (DSA 5000, Anton Paar). The temperature of the apparatus is controlled by Peltier modules and a Pt-100 resistance thermometer, achieving a stability of 0.001 K. Further details on the technique are described elsewhere [39,40].

The calibration of the DSA 5000 was performed by fitting the density values, $\rho$, of reference liquids, obtained from the literature, to the well-established characteristic equation of vibrating-tube densimeters:

$$\rho = A + B\,Q^2 \qquad (2)$$

where $Q$ denotes the ratio between the resonance period of the measuring tube and that of the reference tube, and parameters $A$ and $B$ are determined at each temperature at which the mixtures are to be measured, using the following series of pure liquids: *n*-heptane, isooctane, cyclohexane, toluene, and Milli-Q water. Viscosity corrections have been neglected, since the viscosities of the liquids used are low. The estimated contribution to the uncertainty of density from the calibration parameters $A$ and $B$ is less than 0.00010 g cm$^{-3}$. However, the overall measurement uncertainty of the densimeter is higher and is evaluated as detailed in section 2.4.

The possible contamination and stability of the pure liquids used in this work were monitored periodically by measuring their densities, which remained within a maximum deviation of 2·10⁻⁵ g cm⁻³ from the initial values. The liquid mixtures were prepared by weighing, using an analytical balance (MSU125P, Sartorius), and applying the corresponding correction for buoyancy effects. This procedure resulted in a standard uncertainty of 5·10⁻⁵ g. The calculated concentration is expressed as the mole fraction of 2-propanol, $x_1$, throughout the remainder of the text, unless otherwise indicated.

Calorimetric data were obtained by means of a standard Tian-Calvet microcalorimeter with a temperature stability of 0.01 K. The calorimetric block consists of two wells equipped with thermopiles connected in a differential configuration. One well contains the liquid sample under study (L), while the other holds the reference sample (R).

The instrument was calibrated electrically by supplying a constant power to the sample well until the calorimeter reached a steady state, while measuring the differential electromotive force (emf) at the microcalorimeter output. The calibration constant, $C$, was determined using the following equation:

$$C = \frac{\frac{\delta Q_\text{L}}{\text{d}t} - \frac{\delta Q_\text{R}}{\text{d}t}}{\varepsilon_\text{diff}} = \frac{P}{\varepsilon_\text{diff}} \qquad (3)$$

where $\frac{\delta Q_\text{L}}{\text{d}t} - \frac{\delta Q_\text{R}}{\text{d}t}$ represents the difference between the heat flow from the sample and reference wells, $P$ is the power generated in the sample well (calculated as the product of the voltage and current applied to it), and $\varepsilon_\text{diff}$ is the differential emf.

The calibration setup consists of: (i) Two metallic calibration cells, each containing a wound manganin wire resistor. A power supply (6644A, Agilent) controls both the power supplied to the calibration cell located in the sample well and the duration of the power supply. (ii) A thermal power monitoring module, comprising a multimeter (PM2534, Philips) for measuring the voltage across the resistor, and a multimeter (34410A, Agilent) for measuring the current through the resistor. (iii) A multimeter (2000, Keithley) for monitoring the output voltage generated by the thermopiles; this device is, in fact, the only one required to determine $H_\text{m}^\text{E}$ once the calorimeter has been calibrated.

The calibration procedure is carried out at different supply currents ranging from 5 mA to 35 mA, maintaining each current level stable for approximately 4 hours. The estimated relative standard uncertainty in the calibration constant $C$ is 0.1 %. However, the overall measurement uncertainty increases by an order of magnitude due to factors such as calorimeter signal integration. The total uncertainty in $H_\text{m}^\text{E}$ is determined as described in section 2.4.

The measurement of the $H_\text{m}^\text{E}$ of the two liquids was conducted as follows. Custom-made stainless steel mixing cells, with a total volume of approximately 9 cm³, were used. Prior to mixing, one liquid was placed in the cell, and the other one in a syringe, both housed within the calorimetric block, until thermal equilibrium was reached. The $\varepsilon_\text{diff}$ was then recorded for 10 minutes (baseline) and averaged. Subsequently, the liquid in the syringe was slowly and continuously injected into the cell, producing a signal peak due to mixing. After thermal equilibrium was re-established, the baseline was recorded for 10 minutes and averaged to check that this baseline has not significantly changed and that it is stable enough to ensure that no other thermal effects are influencing the signal. The total heat released by the mixture was determined by multiplying the calibration constant $C$ by the integral of $\varepsilon_\text{diff}$ over the time interval during which the mixture absorbed or released heat. The reference level for the integration in the $\varepsilon_\text{diff}$ vs. time representation was taken as a straight line connecting the initial and final baselines. Further details on the technique are provided in previous works [38,39].

## 2.3. Excess functions

The values $F^{\text{id}}$ of the thermodynamic properties of an ideal mixture, under the same temperature and pressure conditions as the real mixture, are calculated using the Benson and Kiyohara formulas [40–42]:

$$F^{\text{id}} = x_1 F_1 + x_2 F_2 \qquad (F = V_{\text{m}}, H_{\text{m}}, (\partial V_{\text{m}}^{\text{E}}/\partial T)_{p,x}) \qquad (4)$$

$$F^{\text{id}} = \phi_1 F_1 + \phi_2 F_2 \qquad (F = \alpha_p, \kappa_T) \qquad (5)$$

where $F_i$ is the value of the property $F$ for pure component $i$, $x_i$ represents the mole fraction of component $i$, and $\phi_i = x_i V_{\text{m},i}/V_{\text{m}}^{\text{id}}$ is the volume fraction of component $i$. The excess properties $F^{\text{E}}$ are defined as:

$$F^{\text{E}} = F - F^{\text{id}} \qquad (6)$$

In the case of $F = H_{\text{m}}$, the excess molar enthalpy $H_{\text{m}}^{\text{E}}$ is equal to the molar heat of mixing at constant pressure, measured directly by calorimetry.

Interactional effects upon mixing are better evaluated by means of the isochoric excess molar internal energy, $U_{\text{m},V}^{\text{E}}$ [43]:

$$U_{\text{m},V}^{\text{E}} = H_{\text{m}}^{\text{E}} - T \frac{\alpha_p}{\kappa_T} V_{\text{m}}^{\text{E}} \qquad (7)$$

In equation (7), the term $T(\alpha_p/\kappa_T)V_{\text{m}}^{\text{E}}$ is referred to as the equation-of-state term (EoS term), and it is related to the volume variation upon mixing. Due to the lack of $\kappa_T$ values for the mixtures studied in this work, $U_{\text{m},V}^{\text{E}}$ has been estimated by assuming ideal behavior for this quantity, using literature $\kappa_T$ values (see Table 2) and equation (5).

## 2.4. Experimental uncertainty

Details regarding the uncertainty of some of the measurements have already been published in a previous work [44], but they are fully specified below.

***Uncertainty under repeatability conditions.*** This term refers to the uncertainty obtained from the sample standard deviation of measurements performed while maintaining the same procedure, operator, measuring system, operating conditions, location, and source liquids over short time intervals and includes the contribution from calibration constants. It is denoted (with a coverage factor of 2) as $U_1$ [45]. The highest values found in this study are: $U_1(T)$ = 0.02 K, $U_1(p)$ = 10 kPa, $U_1(x_1)$ = 0.00010, $U_1(\rho)$ = 0.00010 g cm$^{-3}$, $U_1(\alpha_p)$ = 0.020·10$^{-3}$ K$^{-1}$, $U_1(H_{\text{m}}^{\text{E}})$ = 0.010·$|H_{\text{m}}^{\text{E}}|$.

***Uncertainty under reproducibility conditions.*** This is based on the expression of the root-mean-square deviation, $\sigma(F)$, of $N_{\text{lit}}$ literature (lit) data points from our experimental (exp) values:

$$\sigma(F) = \left[ \frac{1}{N_{\text{lit}} - 1} \sum_{F_{\text{lit}}} (F_{\text{lit}} - F_{\text{exp}})^2 \right]^{1/2} \qquad (F = \rho, \alpha_p) \qquad (8)$$

The summation is performed over each literature data point $F_{\text{lit}}$ at $T$ = 298.15 K. For pure liquids, the expanded (with a coverage factor of 2) uncertainty evaluated under reproducibility conditions, $U_2$, is estimated as $U_2(F) = 2\,\sigma(F)$. For mixtures, $U_2(F)$ is taken as the maximum value between those corresponding of the pure liquids involved (Table 2).

***Total expanded (with a coverage factor of 2) uncertainty, U***. Ideally, this should include other factors, such as the purity of the compounds used. However, these are very difficult to quantify. Therefore, for the mole fraction of the mixtures, the total expanded uncertainty has been roughly estimated as $U(x_1) = 0.0010$. For $F = \rho, \alpha_p$, the total expanded uncertainty can be calculated as:

$$U(F) = [U_1(F)^2 + U_2(F)^2]^{1/2} \qquad (F = \rho, \alpha_p) \tag{9}$$

Specifically, for the mixtures of 2-propanol with monoglyme, diglyme, triglyme, and tetraglyme, the uncertainty in density, $U(\rho)$, is 0.00097 g cm$^{-3}$, 0.00065 g cm$^{-3}$, 0.0010 g cm$^{-3}$, and 0.0013 g cm$^{-3}$, respectively.

The uncertainties of the excess functions are considerably less sensitive to the purity of the compounds and are typically estimated by comparison with reference values from test systems. The following expressions are used: $U(V_m^E) = 0.010\,|V_m^E|_{max} + 0.005$ cm$^3$ mol$^{-1}$, $U_r(H_m^E) = 0.015$, and $U\left[(\partial V_m^E/\partial T)_{p,x}\right] = 0.010\,|(\partial V_m^E/\partial T)_{p,x}|_{max} + 0.0002$ cm$^3$ mol$^{-1}$ K$^{-1}$.

## 3. Results

The experimental $\rho$ and $V_m^E$ values for (2-propanol + linear polyether) liquid mixtures at $p$ = 0.1 MPa and $T$ = (293.15 to 303.15) K are included in Table 3 as functions of $x_1$. The values of $(\partial V_m^E/\partial T)_{p,x}$ at $p$ = 0.1 MPa and $T$ = 298.15 K are also included in Table 3, calculated as the slope of linear regressions of $V_m^E$ as a function of $T$ in the range $T$ = (293.15 to 303.15) K. The experimental values of $H_m^E$ at $p$ = 0.1 MPa and $T$ = 298.15 K are collected in Table 4.

Some excess properties, $F^E = V_m^E, H_m^E, (\partial V_m^E/\partial T)_{p,x}$, are fitted to Redlich-Kister polynomials [46]:

$$F^E = x_1(1-x_1)\sum_{i=0}^{k-1} A_i(2x_1 - 1)^i \qquad (F^E = V_m^E, H_m^E, (\partial V_m^E/\partial T)_{p,x}) \tag{10}$$

using unweighted linear least-squares regressions, selecting an appropriate number, $k$, of $A_i$ coefficients based on F-tests for the inclusion of additional terms [47]. The results are presented in Table 5. The root-mean-square deviation of the fit to equation (10), $s$, is defined as:

$$s = \left[\frac{1}{N - N_p}\sum_{j=1}^{N}(F_{cal,j} - F_{exp,j})^2\right]^{1/2} \qquad (F = V_m^E, H_m^E, (\partial V_m^E/\partial T)_{p,x}) \tag{11}$$

where $N_p$ denotes the number of $A_i$ parameters in equation (10), and $j$ indexes the number $N$ of experimental data points $F_{exp,j}$ and the corresponding calculated values $F_{cal,j}$. The value of $s$ for each fit is included in Table 5.

The excess properties $V_m^E$, $H_m^E$, and $(\partial V_m^E/\partial T)_{p,x}$ of the studied mixtures at $p$ = 0.1 MPa and $T$ = 298.15 K are depicted as functions of $x_1$ in Figures 1, 2, and 3, respectively, together with the corresponding Redlich-Kister regressions.

The interested reader may find in the Supplementary Material the excess partial molar volumes and enthalpies of both components of the (2-propanol + glyme) liquid mixtures studied (Tables S1 and S2, Figures S1-S4), as well as the corresponding partial molar volumes (Table S1), at $T$ = 298.15 K.

The $V_m^E$ values obtained in this work have been compared with those reported in previous studies at atmospheric pressure [48–51], as illustrated in Figure S5. Figure S5a displays the $V_m^E$ values for the (2-propanol + monoglyme) mixture at temperatures of 293.15 K and 303.15 K. Our data are consistent, within the experimental uncertainty, with those reported by [48] at both temperatures. In contrast, a significant discrepancy is observed when compared with the results of [51], particularly considering that their reported $V_m^E$ values exhibit an inverse temperature dependence—decreasing with increasing temperature—which contradicts both our findings and those of [48].

Figure S5b presents the $V_m^E$ values for the (2-propanol + tetraglyme) mixture at $T$ = 298.15 K. As shown, the density data reported by [49,50] lack the required precision to accurately determine the excess molar volumes, as they yield inconsistent values that fluctuate between positive and negative, and fail to exhibit the expected smooth compositional dependence.

**Table 3.** Density, $\rho$, excess molar volume, $V_m^E$, and temperature derivative of the excess molar volume, $(\partial V_m^E/\partial T)_{p,x}$ of (2-propanol (1) + linear polyether (2)) liquid mixtures as function of the mole fraction of 2-propanol, $x_1$, at temperature $T$ and pressure $p$ = 0.1 MPa.[a]

| $x_1$ | $\rho$/(g cm$^{-3}$) | | | $V_m^E$/(cm$^3$ mol$^{-1}$) | | | $(\partial V_m^E/\partial T)_{p,x}$ / (cm$^3$ mol$^{-1}$ K$^{-1}$) |
|---|---|---|---|---|---|---|---|
| | $T$ = 293.15 K | $T$ = 298.15 K | $T$ = 303.15 K | $T$ = 293.15 K | $T$ = 298.15 K | $T$ = 303.15 K | $T$ = 298.15 K |
| | | | 2-propanol (1) + monoglyme (2) | | | | |
| 0.05178 | 0.86380 | 0.85836 | 0.85291 | 0.0231 | 0.0260 | 0.0276 | 0.00045 |
| 0.10439 | 0.86021 | 0.85481 | 0.84938 | 0.0548 | 0.0598 | 0.0658 | 0.00110 |
| 0.15427 | 0.85678 | 0.85141 | 0.84602 | 0.0771 | 0.0850 | 0.0925 | 0.00154 |
| 0.20588 | 0.85310 | 0.84777 | 0.84241 | 0.1039 | 0.1138 | 0.1246 | 0.00207 |
| 0.30275 | 0.84608 | 0.84084 | 0.83556 | 0.1344 | 0.1466 | 0.1605 | 0.00261 |
| 0.40779 | 0.83815 | 0.83301 | 0.82784 | 0.1531 | 0.1678 | 0.1828 | 0.00297 |
| 0.49539 | 0.83121 | 0.82617 | 0.82109 | 0.1626 | 0.1780 | 0.1945 | 0.00319 |
| 0.60142 | 0.82245 | 0.81755 | 0.81261 | 0.1586 | 0.1732 | 0.1884 | 0.00298 |
| 0.70143 | 0.81382 | 0.80907 | 0.80426 | 0.1357 | 0.1482 | 0.1630 | 0.00273 |
| 0.79875 | 0.80491 | 0.80031 | 0.79567 | 0.1090 | 0.1196 | 0.1302 | 0.00212 |
| 0.85010 | 0.80008 | 0.79558 | 0.79104 | 0.0837 | 0.0915 | 0.0989 | 0.00152 |
| 0.89810 | 0.79539 | 0.79098 | 0.78652 | 0.0619 | 0.0676 | 0.0737 | 0.00118 |
| 0.95061 | 0.79012 | 0.78582 | 0.78148 | 0.0336 | 0.0360 | 0.0377 | 0.00041 |
| | | | 2-propanol (1) + diglyme (2) | | | | |
| 0.05109 | 0.93918 | 0.93422 | 0.92926 | 0.0075 | 0.0107 | 0.0108 | 0.00033 |
| 0.10475 | 0.93418 | 0.92922 | 0.92426 | 0.0170 | 0.0237 | 0.0272 | 0.00102 |
| 0.15607 | 0.92913 | 0.92419 | 0.91925 | 0.0289 | 0.0360 | 0.0399 | 0.00110 |
| 0.19889 | 0.92466 | 0.91973 | 0.91479 | 0.0466 | 0.0552 | 0.0621 | 0.00155 |
| 0.29787 | 0.91373 | 0.90883 | 0.90391 | 0.0630 | 0.0744 | 0.0852 | 0.00222 |
| 0.39453 | 0.90186 | 0.89700 | 0.89211 | 0.0819 | 0.0951 | 0.1087 | 0.00268 |
| 0.49704 | 0.88786 | 0.88305 | 0.87822 | 0.0915 | 0.1063 | 0.1199 | 0.00284 |
| 0.59926 | 0.87218 | 0.86744 | 0.86267 | 0.0893 | 0.1040 | 0.1185 | 0.00292 |
| 0.70252 | 0.85426 | 0.84960 | 0.84492 | 0.0712 | 0.0858 | 0.0986 | 0.00274 |
| 0.79283 | 0.83628 | 0.83173 | 0.82714 | 0.0655 | 0.0769 | 0.0884 | 0.00229 |
| 0.84771 | 0.82438 | 0.81991 | 0.81539 | 0.0379 | 0.0465 | 0.0560 | 0.00181 |
| 0.89995 | 0.81188 | 0.80750 | 0.80306 | 0.0329 | 0.0382 | 0.0451 | 0.00122 |
| 0.94705 | 0.79980 | 0.79550 | 0.79114 | 0.0175 | 0.0204 | 0.0247 | 0.00072 |
| | | | 2-propanol (1) + triglyme (2) | | | | |
| 0.10260 | 0.97560 | 0.97085 | 0.96610 | 0.0057 | 0.0094 | 0.0127 | 0.00070 |
| 0.16302 | 0.96960 | 0.96485 | 0.96010 | 0.0069 | 0.0129 | 0.0181 | 0.00112 |
| 0.21067 | 0.96451 | 0.95977 | 0.95502 | 0.0116 | 0.0177 | 0.0246 | 0.00130 |
| 0.30289 | 0.95374 | 0.94901 | 0.94427 | 0.0138 | 0.0221 | 0.0306 | 0.00168 |
| 0.40500 | 0.94011 | 0.93540 | 0.93068 | 0.0105 | 0.0202 | 0.0297 | 0.00192 |
| 0.49568 | 0.92607 | 0.92138 | 0.91668 | 0.0145 | 0.0259 | 0.0364 | 0.00219 |
| 0.59697 | 0.90781 | 0.90315 | 0.89849 | 0.0097 | 0.0225 | 0.0327 | 0.00230 |
| 0.68977 | 0.88795 | 0.88335 | 0.87874 | 0.0074 | 0.0186 | 0.0279 | 0.00205 |
| 0.79723 | 0.86004 | 0.85553 | 0.85100 | 0.0037 | 0.0128 | 0.0204 | 0.00167 |
| 0.84925 | 0.84410 | 0.83965 | 0.83517 | 0.0009 | 0.0081 | 0.0146 | 0.00137 |
| 0.89995 | 0.82664 | 0.82226 | 0.81785 | −0.0010 | 0.0041 | 0.0081 | 0.00091 |
| 2-propanol (1) + tetraglyme (2) | | | | | | | |
| 0.05776 | 1.00625 | 1.00162 | 0.99699 | −0.0146 | −0.0157 | −0.0129 | 0.00017 |
| 0.10331 | 1.00227 | 0.99763 | 0.99300 | −0.0164 | −0.0144 | −0.0109 | 0.00055 |
| 0.16636 | 0.99631 | 0.99168 | 0.98705 | −0.0162 | −0.0148 | −0.0104 | 0.00058 |
| 0.19980 | 0.99294 | 0.98830 | 0.98368 | −0.0199 | −0.0157 | −0.0128 | 0.00071 |
| 0.29568 | 0.98226 | 0.97763 | 0.97300 | −0.0291 | −0.0245 | −0.0181 | 0.00110 |
| 0.40048 | 0.96853 | 0.96390 | 0.95928 | −0.0412 | −0.0337 | −0.0271 | 0.00141 |
| 0.49985 | 0.95293 | 0.94832 | 0.94371 | −0.0513 | −0.0439 | −0.0366 | 0.00147 |
| 0.59618 | 0.93466 | 0.93008 | 0.92549 | −0.0628 | −0.0561 | −0.0489 | 0.00139 |
| 0.69890 | 0.91048 | 0.90594 | 0.90139 | −0.0688 | −0.0623 | −0.0565 | 0.00123 |
| 0.79196 | 0.88272 | 0.87824 | 0.87374 | −0.0658 | −0.0606 | −0.0558 | 0.00100 |
| 0.84810 | 0.86227 | 0.85784 | 0.85339 | −0.0573 | −0.0530 | −0.0499 | 0.00074 |
| 0.90240 | 0.83902 | 0.83466 | 0.83027 | −0.0474 | −0.0453 | −0.0438 | 0.00036 |
| 0.94957 | 0.81522 | 0.81093 | 0.80660 | −0.0287 | −0.0277 | −0.0271 | 0.00016 |

[a] Total expanded uncertainties ($U$), with a coverage factor of 2: $U(T)$ = 0.02 K, $U(p)$ = 10 kPa, $U(x_1)$ = 0.0010; $U(\rho)$ is estimated as $U_1(\rho)$ = 0.00010 g cm$^{-3}$ plus the maximum $U_2(\rho) = 2\sigma(\rho)$ value between those of the pure liquids involved (see Table 2), that is, $U(\rho) = [U_1(\rho)^2 + U_2(\rho)^2]^{1/2}$; $U(V_m^E)$ = 0.010 $|V_m^E|_{max}$ + 0.005 cm$^3$ mol$^{-1}$; $U((\partial V_m^E/\partial T)_{p,x})$ = 0.010 $|(\partial V_m^E/\partial T)_{p,x}|_{max}$ + 0.0002 cm$^3$ mol$^{-1}$ K$^{-1}$.

**Table 4.** Excess molar enthalpy, $H_m^E$, of (2-propanol (1) + linear polyether (2)) liquid mixtures as function of the mole fraction of 2-propanol, $x_1$, at temperature $T$ = 298.15 K and pressure $p$ = 0.1 MPa.[a]

| $x_1$ | $H_m^E$/(J mol$^{-1}$) | $x_1$ | $H_m^E$/(J mol$^{-1}$) | $x_1$ | $H_m^E$/(J mol$^{-1}$) | $x_1$ | $H_m^E$/(J mol$^{-1}$) |
|---|---|---|---|---|---|---|---|
| 2-propanol (1) + monoglyme (2) | | 2-propanol (1) + diglyme (2) | | 2-propanol (1) + triglyme (2) | | 2-propanol (1) + tetraglyme (2) | |
| 0.05116 | 266 | 0.05862 | 310 | 0.04112 | 194 | 0.04079 | 219 |
| 0.10266 | 504 | 0.09944 | 523 | 0.10454 | 580 | 0.10101 | 573 |
| 0.14753 | 681 | 0.14873 | 750 | 0.15061 | 811 | 0.14880 | 839 |
| 0.20743 | 890 | 0.19941 | 962 | 0.19644 | 1020 | 0.20190 | 1123 |
| 0.31511 | 1145 | 0.29935 | 1314 | 0.29930 | 1413 | 0.30358 | 1517 |
| 0.39259 | 1257 | 0.39978 | 1503 | 0.40307 | 1697 | 0.40874 | 1834 |
| 0.50397 | 1324 | 0.50348 | 1626 | 0.50693 | 1858 | 0.49262 | 1996 |
| 0.60170 | 1279 | 0.59184 | 1615 | 0.59909 | 1863 | 0.59730 | 2044 |
| 0.71260 | 1106 | 0.69626 | 1483 | 0.69581 | 1744 | 0.70215 | 1938 |
| 0.79651 | 895 | 0.80243 | 1179 | 0.80344 | 1407 | 0.80673 | 1605 |
| 0.85240 | 697 | 0.85192 | 958 | 0.85000 | 1195 | 0.84681 | 1411 |
| 0.89572 | 523 | 0.89319 | 758 | 0.89311 | 916 | 0.90341 | 1022 |
| 0.93882 | 328 | 0.97297 | 220 | 0.95217 | 477 | 0.95277 | 574 |

[a] Total expanded uncertainties ($U$) and relative expanded uncertainties ($U_r$) with a coverage factor of 2: $U(T)$ = 0.02 K, $U(p)$ = 10 kPa, $U(x_1)$ = 0.0010, $U_r(H_m^E)$ = 0.015.

**Table 5.** Parameters $A_i$ of equation (10) and root-mean-square deviation of the fit, $s$, equation (11), for the excess molar volume, $V_m^E$, the temperature derivative of the excess molar volume, $(\partial V_m^E/\partial T)_{p,x}$, and the excess molar enthalpy, $H_m^E$, of (2-propanol (1) + linear polyether (2)) liquid mixtures used in this work at temperature $T$ and pressure $p$ = 0.1 MPa.

| Property | $T$/K | $A_0$ | $A_1$ | $A_2$ | $A_3$ | $s$ |
|---|---|---|---|---|---|---|
| 2-propanol (1) + monoglyme (2) | | | | | | |
| $V_m^E$/(cm$^3$ mol$^{-1}$) | 293.15 | 0.644 | 0.05 | | | 0.003 |
| | 298.15 | 0.705 | 0.05 | | | 0.003 |
| | 303.15 | 0.769 | 0.05 | | | 0.003 |
| $(\partial V_m^E/\partial T)_{p,x}$/(cm$^3$ mol$^{-1}$ K$^{-1}$) | 298.15 | 0.0125 | | | | 0.00010 |
| $H_m^E$/(J mol$^{-1}$) | 298.15 | 5292 | 97 | 410 | | 3 |
| 2-propanol (1) + diglyme (2) | | | | | | |
| $V_m^E$/(cm$^3$ mol$^{-1}$) | 293.15 | 0.36 | 0.08 | −0.15 | | 0.004 |
| | 298.15 | 0.42 | 0.09 | −0.14 | | 0.004 |
| | 303.15 | 0.48 | 0.10 | −0.16 | | 0.004 |
| $(\partial V_m^E/\partial T)_{p,x}/(\text{cm}^3\,\text{mol}^{-1}\,\text{K}^{-1})$ | 298.15 | 0.0117 | 0.0032 | | | 0.00008 |
| $H_m^E/(\text{J mol}^{-1})$ | 298.15 | 6496 | 1165 | 604 | | 13 |
| 2-propanol (1) + triglyme (2) | | | | | | |
| $V_m^E/(\text{cm}^3\,\text{mol}^{-1})$ | 293.15 | 0.045 | −0.037 | | | 0.0019 |
| | 298.15 | 0.092 | −0.022 | | | 0.0019 |
| | 303.15 | 0.134 | −0.016 | | | 0.0020 |
| $(\partial V_m^E/\partial T)_{p,x}/(\text{cm}^3\,\text{mol}^{-1}\,\text{K}^{-1})$ | 298.15 | 0.0089 | 0.0021 | | | 0.00007 |
| $H_m^E/(\text{J mol}^{-1})$ | 298.15 | 7383 | 1691 | 825 | 992 | 14 |
| 2-propanol (1) + tetraglyme (2) | | | | | | |
| $V_m^E/(\text{cm}^3\,\text{mol}^{-1})$ | 293.15 | −0.204 | −0.23 | −0.20 | | 0.003 |
| | 298.15 | −0.174 | −0.22 | −0.22 | | 0.003 |
| | 303.15 | −0.145 | −0.22 | −0.22 | | 0.003 |
| $(\partial V_m^E/\partial T)_{p,x}/(\text{cm}^3\,\text{mol}^{-1}\,\text{K}^{-1})$ | 298.15 | 0.0055 | | | | 0.00011 |
| $H_m^E/(\text{J mol}^{-1})$ | 298.15 | 7982 | 2233 | 1590 | 1715 | 11 |

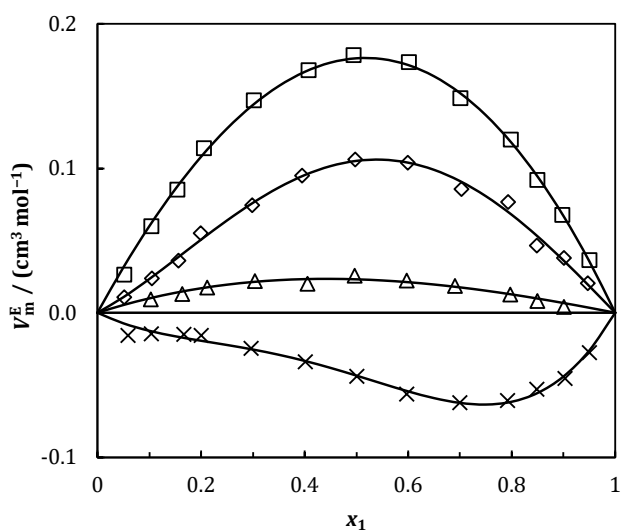

**Fig. 1**. Excess molar volume, $V_m^E$, of (2-propanol (1) + linear polyether (2)) liquid mixtures as a function of the 2-propanol mole fraction, $x_1$, at temperature $T$ = 298.15 K and pressure $p$ = 0.1 MPa. Symbols, experimental values: (□) monoglyme, (◇) diglyme, (△) triglyme, (×) tetraglyme. Solid lines are calculated with equation (10) using the coefficients from Table 5.

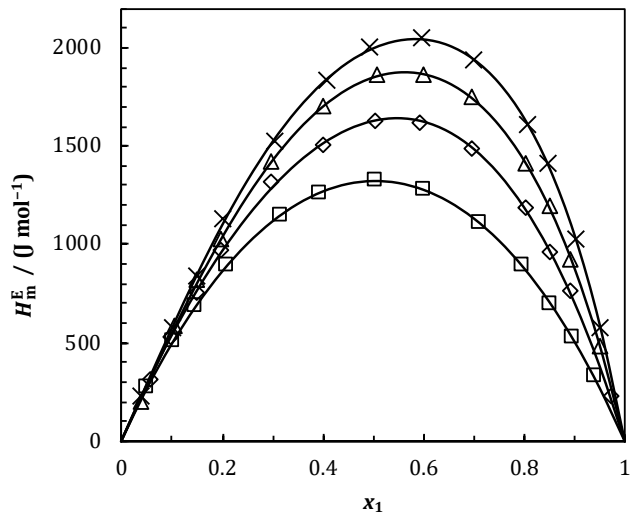

**Fig. 2.** Excess molar enthalpy, $H_m^E$, of (2-propanol (1) + linear polyether (2)) liquid mixtures as a function of the 2-propanol mole fraction, $x_1$, at temperature $T$ = 298.15 K and pressure $p$ = 0.1 MPa. Symbols, experimental values: (□) monoglyme, (◇) diglyme, (△) triglyme, (×) tetraglyme. Solid lines are calculated with equation (10) using the coefficients from Table 5.

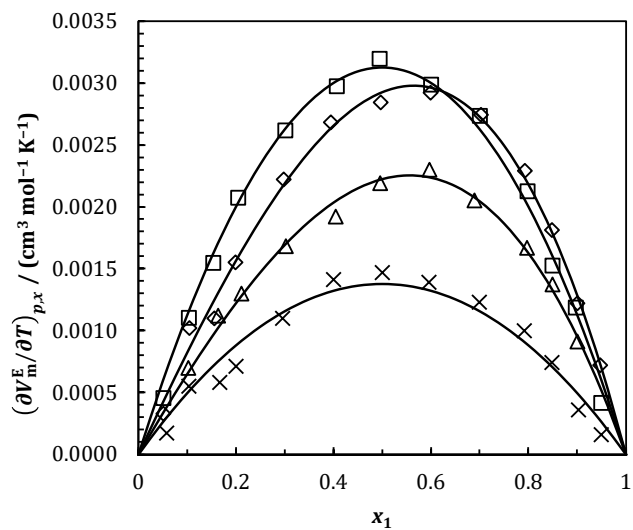

**Fig. 3.** Temperature derivative of the excess molar volume, $(\partial V_m^E/\partial T)_{p,x}$, of (2-propanol (1) + linear polyether (2)) liquid mixtures as a function of the 2-propanol mole fraction, $x_1$, at temperature $T$ = 298.15 K and pressure $p$ = 0.1 MPa. Symbols, experimental values: (□) monoglyme, (◇) diglyme, (△) triglyme, (×) tetraglyme. Solid lines are calculated with equation (10) using the coefficients from Table 5.

## 4. Flory model

The essential hypotheses of the theory [52–54] are described elsewhere [55]. The basic assumption of the model is that of random mixing. The explicit expression of the Flory equation of state, valid for both pure liquids and liquid mixtures, is:

$$\frac{\hat{p}\hat{V}}{\hat{T}} = \frac{\hat{V}^{1/3}}{\hat{V}^{1/3}-1} - \frac{1}{\hat{V}\hat{T}} \tag{12}$$

where $\hat{V} = V_m/V_m^*$, $\hat{p} = p/p^*$, and $\hat{T} = T/T^*$ stand for the reduced volume, pressure and temperature, respectively (where $V_m$ is the molar volume of the mixture). For pure liquids, the reduction parameters, $V_{m,i}^*$, $p_i^*$, and $T_i^*$, can be obtained from values of density, $\alpha_{p,i}$ (isobaric expansion coefficient), and $\kappa_{T,i}$ (isothermal compressibility) (see Table 2). Expressions for the reduction parameters of mixtures can be found elsewhere [55]. The reduction parameters used in this work are listed in Table 6. The values of $H_m^E$ are obtained from:

$$H_m^E = \frac{x_1 V_{m,1}^* \theta_2 X_{12}}{\hat{V}} + x_1 V_{m,1}^* p_1^* \left(\frac{1}{\hat{V}_1} - \frac{1}{\hat{V}}\right) + x_2 V_{m,2}^* p_2^* \left(\frac{1}{\hat{V}_2} - \frac{1}{\hat{V}}\right) \quad (13)$$

where all symbols have their usual meaning [55]. The reduced volume of the mixture, $\hat{V}$, is obtained from the equation of state, and the excess molar volume can be calculated as follows:

$$V_m^E = (x_1 V_{m,1}^* + x_1 V_{m,2}^*)(\hat{V} - \varphi_1 \hat{V}_1 - \varphi_2 \hat{V}_2) \quad (14)$$

## 4.1. Results from Flory's model

Table 6 lists the values of the interaction parameter $X_{12}$, determined from experimental $H_m^E$ data at equimolar composition and $T$ = 298.15 K, using the method described in reference [56]. Deviations between experimental and theoretical $H_m^E$ results (see Table 6 and Figure 4 for a graphical comparison) are evaluated using the relative standard deviations of $H_m^E$, defined as:

$$\sigma_r(H_m^E) = \left[\frac{1}{N}\sum \left(\frac{H_{m,exp}^E - H_{m,calc}^E}{H_{m,exp}^E}\right)^2\right]^{1/2} \quad (15)$$

where $N$ is the number of experimental data points. Finally, experimental $V_m^E$ results at $x_1$ = 0.5 and $T$ = 298.15 K are also compared with the corresponding Flory calculations, as shown in Table 6.

**Table 6.** Results from the Flory model for (2-propanol (1) + ether (2)) mixtures at $p$ = 0.1 MPa and $T$ = 298.15 K: interaction parameter, $X_{12}$, relative standard deviations for $H_m^E$ from equation (13), $\sigma_r(H_m^E)$, and excess molar volumes at equimolar composition, $V_m^E$. Flory parameters of pure compounds[a] at $T$ = 298.15 K are also included.

|  | $V_{m,i}$ /(cm³ mol⁻¹) | $V_{m,i}^*$ /(cm³ mol⁻¹) | $p_i^*$ /(J cm⁻³) | $N$[b] | $X_{12}$ /(J cm⁻³) | $\sigma_r(H_m^E)$ | $V_m^E$ /(cm³ mol⁻¹) Exp. | $V_m^E$ /(cm³ mol⁻¹) Flory |
|---|---|---|---|---|---|---|---|---|
| Monoglyme | 104.58 | 80.39 | 575.9 | 13[c] | 75.53 | 0.051 | 0.176[c] | 1.257 |
| Diglyme | 142.93 | 113.47 | 609.8 | 13[c] | 84.65 | 0.029 | 0.105[c] | 1.140 |
| Triglyme | 181.84 | 146.59 | 626.1 | 13[c] | 90.76 | 0.056 | 0.023[c] | 1.024 |
| Tetraglyme | 220.89 | 179.43 | 645.5 | 13[c] | 94.18 | 0.041 | −0.044[c] | 0.914 |
| Dipropylether | 137.68 | 106 | 440.7 | 14[d] | 50.63 | 0.273 | −0.027[d] | 0.925 |

[a] $V_{m,i}$, molar volume; $V_{m,i}^*$ and $p_i^*$, reduction molar volume and pressure, respectively. For 2-propanol: $V_{m,i}$ = 76.96 cm³ mol⁻¹; $V_{m,i}^*$ = 60.87 cm³ mol⁻¹; $p_i^*$ = 464.5 J cm⁻³.

[b] Number of data points.

[c] This work.

[d] Reference [57].

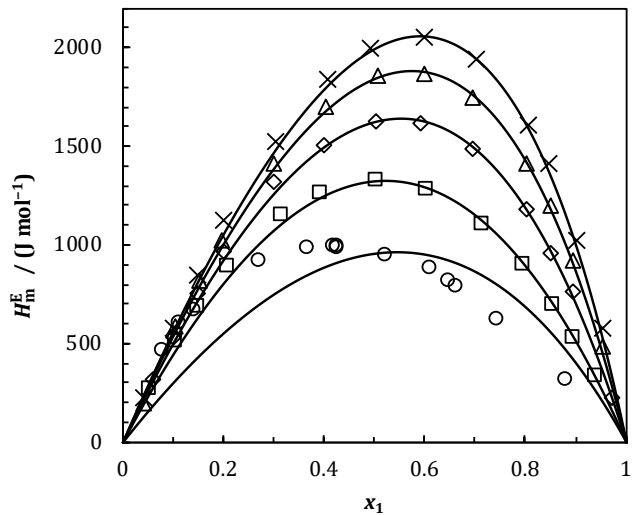

**Fig. 4**. Excess molar enthalpy, $H_m^E$, of (2-propanol (1) + linear polyether (2) or + di-$n$-propyl ether (2)) liquid mixtures as a function of the 2-propanol mole fraction, $x_1$, at temperature $T$ = 298.15 K and pressure $p$ = 0.1 MPa. Symbols, experimental values: (□) monoglyme [this work], (◇) diglyme [this work], (△) triglyme [this work], (×) tetraglyme [this work], (○) di-$n$-propyl ether [57]. Solid lines are calculated with Flory model (equation (13)) using coefficients from Table 6.

## 5. Discussion

Unless otherwise indicated, the values of the thermophysical properties are considered at $T$ = 298.15 K, $p$ = 0.1 MPa, and $x_1 = 0.5$ (see Tables 7 and 8, and Figures 5 and 6) hereinafter. The number of carbon atoms in the $n$-alkane is denoted by $n$, and the number of oxygen atoms in the glyme is denoted by $u$.

**Table 7.** Excess molar volume, $V_m^E$, temperature derivative of excess molar volume, $(\partial V_m^E/\partial T)_{p,x}$, of (1- and 2-propanol + glyme) liquid mixtures at equimolar composition, at temperature $T$ = 298.15 K, and pressure $p$ = 0.1 MPa.

| | 2-propanol | | 1-propanol | |
|---|---|---|---|---|
| Glyme | $V_m^E$/(cm³ mol⁻¹) | $(\partial V_m^E/\partial T)_{p,x}$ /10⁻³(cm³ mol⁻¹ K⁻¹) | $V_m^E$/(cm³ mol⁻¹) | $(\partial V_m^E/\partial T)_{p,x}$ /10⁻³(cm³ mol⁻¹ K⁻¹) |
| monoglyme | 0.176 | 3.13 | −0.071 [58] | 1.23 [59] |
| diglyme | 0.105 | 2.93 | −0.068 [58] | 2.92 [59] |
| triglyme | 0.023 | 2.23 | −0.120 [60] | 2.24 [59] |
| tetraglyme | −0.044 | 1.38 | −0.202 [59] | 1.08 [59] |

**Table 8.** Excess molar enthalpy, $H_m^E$, isochoric excess molar internal energy, $U_{m,V}^E$, and OH−O interaction enthalpy, $\Delta H_{OH-O}$, of (1- and 2-propanol + glyme) liquid mixtures at equimolar composition, at temperature $T$ = 298.15 K and pressure $p$ = 0.1 MPa.

| | 2-propanol | | | 1-propanol | | |
|---|---|---|---|---|---|---|
| Glyme | $H_m^E$/ (J mol⁻¹) | $U_{m,V}^E$ / (J mol⁻¹) | $\Delta H_{OH-O}$/ (kJ mol⁻¹) | $H_m^E$/ (J mol⁻¹) | $U_{m,V}^E$ / (J mol⁻¹) | $\Delta H_{OH-O}$/ (kJ mol⁻¹) |
| monoglyme | 1323 | 1266 | −23.1 | 1040 [61] | 1063 | −23.9 [8] |
| diglyme | 1624 | 1587 | −25.4 | 1214 [62] | 1249 | −26.6 [8] |
| triglyme | 1846 | 1837 | −28.5 | 1401 [63] | 1450 | −30 [8] |
| tetraglyme | 1996 | 2012 | −30.9 | 1514 [64] | 1600 | −34.2 [8] |

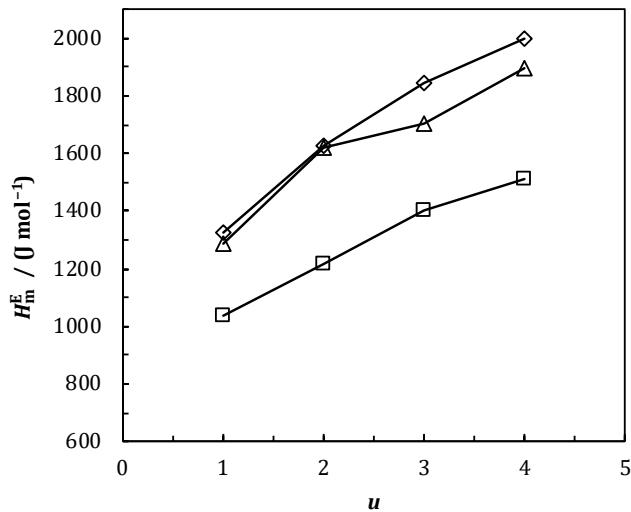

**Fig. 5**. Excess molar enthalpy, $H_\text{m}^\text{E}$, of (glyme + alkane) or (alkanol + glyme) systems as a function of $u$, the number of oxygen atoms in the glyme, at equimolar composition, $p$ = 0.1 MPa and $T$ = 298.15 K: ($\diamond$) 2-propanol [this work]; ($\square$) 1-propanol [61–64]; ($\triangle$) $n$-heptane [65].

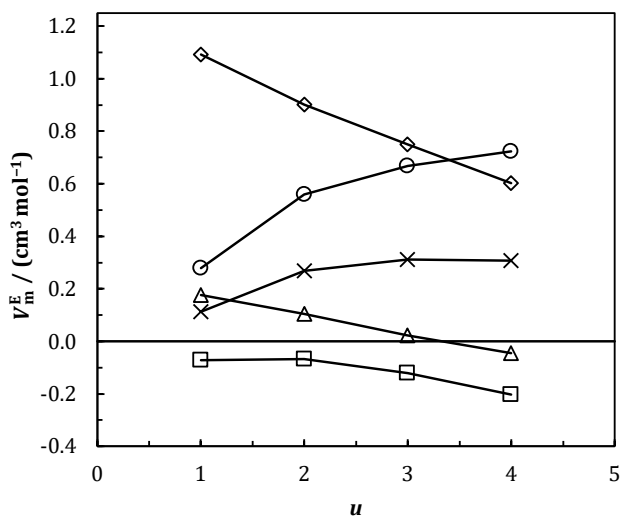

**Fig. 6**. Excess molar volume, $V_\text{m}^\text{E}$, of (glyme + alkane) or (alkanol + glyme) systems as a function of $u$, the number of oxygen atoms in the glyme, at equimolar composition, $p$ = 0.1 MPa and $T$ = 298.15 K: ($\triangle$) 2-propanol [this work]; ($\square$) 1-propanol [58–60]; ($\diamond$) $n$-heptane [34,66,67]; ($\circ$) 1-nonanol [68]; ($\times$) 1-hexanol [69].

## 5.1.  (Alkanol + *n*-alkane) mixtures

Firstly, we focus on systems containing 1-alkanols. For mixtures with a fixed *n*-alkane (e.g., *n*-heptane), increasing the size of the 1-alkanol leads to an interesting behavior in both $H_\text{m}^\text{E}$ and $V_\text{m}^\text{E}$. The (methanol + *n*-heptane) mixture exhibits a miscibility gap between 0.17 and 0.89 mole fraction of methanol at $T$ = 298.15 K [70], due to the difficulty of the *n*-alkane in breaking the strong self-association of this alcohol. This miscibility gap disappears at 298.15 K in *n*-heptane systems involving other 1-alkanols, as the –OH group becomes more sterically hindered, leading to reduced self-association of the alkanol [63]. The $H_\text{m}^\text{E}$ values are positive (arising from the disruption of interactions between alkanol molecules) and increase from ethanol to 1-propanol [71], but then gradually decrease [72]. This behavior is typically explained considering that, due to the lower self-association of longer 1-alkanols, the disruption of their intermolecular network upon mixing contributes less positively to $H_m^E$.

On the other hand, the sign of $V_m^E$ depends on the size difference between the components of the mixture: $V_m^E$ is positive for the shorter 1-alkanols and becomes negative starting from 1-hexanol. These negative values of $V_m^E$ have been explained in terms of a purely structural effect known as interstitial accommodation [73,74].

Fixing 1-propanol as the 1-alkanol, the values of $H_m^E$ for (1-propanol + n-alkane) mixtures increase slightly with the size of the n-alkane: $H_m^E/(\text{J mol}^{-1})$ = 533 (n-hexane [75]) < 614 (n-heptane [76]) < 685 (n-octane [77]), which can be attributed to the greater ability of longer alkanes to break interactions between 1-alkanol molecules. However, these small variations in $H_m^E$ suggest that the self-associative structures of the alkanols are only slightly perturbed by the alkanes. Accordingly, the $H_m^E(x_1)$ curves are skewed toward low mole fractions of the 1-alkanol. The same trend is observed in the corresponding $V_m^E/(\text{cm}^3 \text{ mol}^{-1})$ values: 0.180 (n-hexane [78]) < 0.295 (n-heptane [73]).

Let us now examine the effect of replacing 1-propanol with 2-propanol. As a starting point, it is important to note that the enthalpy of self-association of 2-propanol can be assumed to be equal to that of 1-alkanols. This approximation has been widely used in the application of association theories [63,79–82]. Nevertheless, the self-association of 2-propanol is lower than that of 1-propanol, since its –OH group is more sterically hindered [63]. For mixtures with n-heptane, $H_m^E/(\text{J mol}^{-1})$ = 614 (1-propanol [76]) < 797 (2-propanol) [83]. This suggests that, in the case of 2-propanol mixtures, the breaking of dipolar interactions between alcohol molecules also contributes significantly to $H_m^E$. A similar behavior is observed in the corresponding $V_m^E/(\text{cm}^3 \text{ mol}^{-1})$ values for mixtures with n-hexane: 0.180 (1-propanol [78]) < 0.410 (2-propanol [84]), which can be explained by the same interactional effect.

On the other hand, the high and positive values of the isobaric molar excess heat capacities, $C_{p,m}^E/(\text{J mol}^{-1} \text{ K}^{-1})$, for mixtures with n-heptane: ≈13.5 (1-propanol [85]) and 15.2 (2-propanol [86]), highlight that, in both solutions, the alcohol network is significantly disrupted as the temperature increases. However, the effect is more pronounced when 2-propanol is involved.

## 5.2. (Glyme + n-alkane) mixtures

Glymes are linear polyethers containing multiple –O– groups within the molecule. They are characterized by both inter and intramolecular dipolar interactions [87]. When glymes are mixed with n-alkanes, these dipolar interactions between glyme molecules are partially disrupted.

For a fixed glyme, $H_m^E$ of (glyme + alkane) mixtures are positive, considerably high and increase with the chain length of the n-alkane. For instance, the $H_m^E/(\text{J mol}^{-1})$ values for (triglyme + n-alkane) are [88]: 1704 (for n-heptane) < 1877 (for n-octane) < 2110 (for n-decane) < 2214 (for n-dodecane). This behavior is a consequence of the intense dipolar interactions between glyme molecules, which are more effectively disrupted by longer alkanes. Interestingly, the (tetraglyme + n-dodecane) mixture exhibits a miscibility gap at 298.15 K in the concentration range $0.20 < x_1 < 0.74$ [89]. The observed immiscibility reflects the inability of n-dodecane to sufficiently disrupt the intense dipolar interactions between molecules of tetraglyme. $V_m^E$ of (glyme + $n$-alkane) systems for a fixed glyme behaves in a similar way. For example, in (triglyme + $n$-alkane) solutions, the $V_m^E/(\text{cm}^3\,\text{mol}^{-1})$ values are: 0.748 ($n$-heptane [34]) < 0.970 ($n$-octane [34]) < 1.272 ($n$-decane [34]) < 1.478 ($n$-dodecane [90]).

For a fixed $n$-alkane ($n$-heptane), $H_m^E/(\text{J mol}^{-1})$ of (glyme + $n$-heptane) mixtures increases with the size of the glyme molecule: 1285 (monoglyme [65]) < 1621 (diglyme [91]) < 1704 (triglyme [88]) < 1897 (tetraglyme [92]) (Figure 5). This can be attributed to the disruption of stronger dipolar interactions by $n$-heptane in longer glymes, which arise from increasing proximity effects (intramolecular effects) between the oxygen atoms in the glymes [87]. This behavior is also reflected in the values of the partial molar excess enthalpy of glyme at infinite dilution, $\bar{H}_{m,1}^{E,\infty}$, in (glyme + $n$-heptane) mixtures (see section 5.4. below). However, structural effects are also relevant in these mixtures, since $V_m^E/(\text{cm}^3\,\text{mol}^{-1})$ varies in the opposite direction: 1.092 (monoglyme [67]) > 0.902 (diglyme [66]) > 0.749 (triglyme [34]) > 0.602 (tetraglyme [34]) (Figure 6). This quantity may even become negative in solutions with components of very different size, such as (tetraglyme + $n$-pentane), for which $V_m^E = -0.39\,\text{cm}^3\,\text{mol}^{-1}$ [93].

### 5.3. (Alkanol + glyme) mixtures

In (2-propanol + glyme) mixtures, $H_m^E$ values are large and positive (Table 8, Figure 5). This can be attributed to a dominant contribution from the breakdown of interactions between molecules of the same chemical species. Notably, for systems with a given glyme, $H_m^E$ follows the sequence: (2-propanol + glyme) > (glyme + $n$-heptane) > (2-propanol + $n$-heptane) (see Figure 5). This clearly indicates that self-association and solvation effects are of minor importance in (2-propanol + glyme) mixtures, which are largely unstructured. For the (2-propanol + di-$n$-propyl ether, DPE) system, $H_m^E/(\text{J mol}^{-1})$ = 958 [94], a value close to that of the (2-propanol + $n$-heptane) system (797 J mol$^{-1}$ [83]). This suggests that alcohol self-association effects remain relevant in the latter mixture.

For longer glymes, $H_m^E$ is higher, and the maxima of the corresponding $H_m^E(x_1)$ curves are shifted toward higher concentrations of 2-propanol (Figure 2). This can be interpreted as a consequence of the increased ability of longer polyether chains to break (2-propanol)-(2-propanol) interactions. The corresponding values of the isochoric excess molar internal energy, $U_{m,V}^E$ (Table 8), which increase more rapidly than the $H_m^E$ values with the glyme size, support this interpretation. It should be noted that the equation of state term is rather small for these systems, so the $H_m^E$ values are mainly determined by interactional effects. Note that, for the (2-propanol + DPE) system, the $H_m^E(x_1)$ curve is skewed to lower mole fractions of the alkanol (Figure 4), further suggesting that self-association effects are still significant in this mixture.

In contrast with these $H_m^E$ values, the $V_m^E$ values for (2-propanol + glyme) solutions are either small and positive (monoglyme, diglyme and triglyme) or negative (tetraglyme) (see below). Systems exhibiting such values of $V_m^E$, along with largely positive $H_m^E$ results, are characterized by strong structural effects that outweigh the positive interactional contribution to $V_m^E$ arising from the rupture of interactions between like molecules. The relative influence of structural effects on $V_m^E/(\text{cm}^3\,\text{mol}^{-1})$ appears to increase with glyme size, as this excess function becomes progressively lower: 0.176 (monoglyme) > 0.105 (diglyme) > 0.023 (triglyme) > −0.044 (tetraglyme) (Table 7, Figure 6). Note that these variations are the same as those for (glyme + $n$-heptane) mixtures (see section 5.2 above).

The values of $\left(\partial V_m^E/\partial T\right)_{p,x}$ (Table 7) are positive, indicating that the interactional contribution to $V_m^E$ increases with temperature. However, $\left(\partial V_m^E/\partial T\right)_{p,x}$ decreases as glyme size increases (Table 7), which may be due to the growing relevance of structural effects under these conditions.

(1-Propanol + glyme) mixtures are characterized by analogous properties to those of (2-propanol + glyme) solutions. However, for a fixed glyme, replacing 2-propanol with 1-propanol leads to lower values of both $H_m^E$ and $V_m^E$ (Tables 7 and 8, Figures 5 and 6). This behavior can be explained, at least partially, by the combination of the following factors: (i) Glyme molecules break the hydrogen bonding network of 2-propanol more effectively than that of the 1-propanol. This phenomenon is also observed in (alkanol + $n$-heptane) mixtures (see section 5.1 above). (ii) Alkanol-glyme interactions are likely stronger in the case of 1-propanol, as its –OH group is less sterically hindered, resulting in a more negative contribution to the excess functions. (iii) For the same reason, the number of unlike molecules interactions formed upon mixing is greater in these solutions. It is worth noting that in systems such as (1-propanol or 2-propanol + glyme) and (glyme + $n$-heptane), both excess functions, $H_m^E$ and $V_m^E$, vary similarly with glyme size (Figures 5 and 6). This highlights the relevance of both physical interactions and structural effects in the mixtures of alkanols studied. The larger values of $\left(\partial V_m^E/\partial T\right)_{p,x}$ observed for (1-propanol + glyme) systems (Table 7) may be explained by the fact that glymes more effectively disrupt the self-association of 2-propanol.

## 5.4. Analysis of different contributions to excess molar enthalpy

Next, the enthalpy of interaction between 2-propanol and glymes, denoted as $\Delta H_{\text{OH}-\text{O}}$, is evaluated using the equation [8,95]:

$$\Delta H_{\text{OH}-\text{O}} = \bar{H}_{m,1}^{E,\infty} \text{(2-propanol + glyme)} \\ -\bar{H}_{m,1}^{E,\infty}\text{(2-propanol + }n\text{-heptane)} \\ -\bar{H}_{m,1}^{E,\infty}\text{(glyme + }n\text{-heptane)} \quad (16)$$

where $\bar{H}_{m,1}^{E,\infty}$ represents the partial excess molar enthalpy of the first component of the mixture at infinite dilution. In the calculation of $\Delta H_{\text{OH}-\text{O}}$, several considerations have been taken into account:

(i) The values of $\bar{H}_{m,1}^{E,\infty}$ for (2-propanol + glyme) mixtures are obtained from our measurements of $H_m^E$ over the entire mole fraction range. The Redlich-Kister fit is used to calculate these values from the following expression:

$$\bar{H}_{m,1}^{E,\infty} = \lim_{x_1 \to 0}\left[H_m^E + x_2\left(\frac{\partial H_m^E}{\partial x_1}\right)_{T,p}\right] = \sum_{i=0}^{k-1}(-1)^i A_i \quad (17)$$

where the parameters $A_i$ are listed in Table 5. Accordingly, the values of $\bar{H}_{m,1}^{E,\infty}/\left(\text{kJ mol}^{-1}\right)$ are: 5.6 (monoglyme); 5.9 (diglyme); 5.5 (triglyme); 5.6 (tetraglyme).

(ii) As is customary in the application of association theories, it is assumed that $\bar{H}_{m,1}^{E,\infty}$(2-propanol + $n$-heptane) is independent of the alcohol [63,79–82]. In this work, we have used $\bar{H}_{m,1}^{E,\infty}$(2-propanol + $n$-heptane) = 23.2 kJ mol$^{-1}$, which is the same value reported for mixtures with 1-alkanols [96–98].

(iii) Finally, the values of $\bar{H}_{\mathrm{m,1}}^{\mathrm{E},\infty}/(\mathrm{kJ\ mol}^{-1})$ for (glyme + *n*-heptane) mixtures are: 5.5 (monoglyme) [65]; 8.2 (diglyme) [91]; 10.8 (triglyme) [88]; 13.3 (tetraglyme) [92]. These values are consistent with the statement made above regarding the stronger dipolar interactions exhibited by longer glymes (see section 5.2).

An inspection of the $\Delta H_{\mathrm{OH-O}}$ results, collected in Table 8, reveals several interesting findings: (i) Increasing the glyme size in mixtures with a fixed alkanol (1-propanol or 2-propanol) leads to more negative $\Delta H_{\mathrm{OH-O}}$ values, indicating stronger interactions between unlike molecules. This is in accordance with the stronger polar character of longer glymes, as reflected in the $\bar{H}_{\mathrm{m,1}}^{\mathrm{E},\infty}$(glyme + *n*-heptane) values. (ii) Although the $\Delta H_{\mathrm{OH-O}}$ values are similar for both alkanols, 1-propanol consistently shows a higher contribution to interactions between unlike molecules, especially in mixtures with tetraglyme. This can be attributed, as previously discussed, to the fact that the –OH group in 2-propanol is more sterically hindered, reducing its ability to participate in interactions between unlike molecules. (iii) The large and negative $\Delta H_{\mathrm{OH-O}}$ values contrast with the large and positive $H_{\mathrm{m}}^{\mathrm{E}}$ results, indicating that the latter are predominantly determined by the breaking of interactions between like molecules.

We have also determined the $\Delta H_{\mathrm{OH-O}}$ values for the (1-propanol or 2-propanol + DPE) systems. For this purpose, we used the following values of $\bar{H}_{\mathrm{m,1}}^{\mathrm{E},\infty}/(\mathrm{kJ\ mol}^{-1})$: 0.8 (DPE + *n*-heptane) [99]; 6.6 (1-propanol + DPE); 7.5 (2-propanol + DPE) [57]. Thus, the resulting $\Delta H_{\mathrm{OH-O}}/(\mathrm{kJ\ mol}^{-1})$ values are: –15.8 (1-propanol + DPE); –14.9 (2-propanol + DPE). These values are less negative than those obtained for the corresponding solutions with glymes, indicating that interactions between unlike molecules are stronger in the latter mixtures. This is likely because the number of oxygen atoms is higher in polyethers and the –O– group is very sterically hindered in DPE. The lower values of $H_{\mathrm{m}}^{\mathrm{E}}/(\mathrm{J\ mol}^{-1})$ for the systems with DPE (740 (1-propanol); 958 (2-propanol)) [57] suggest that this ether is not an effective breaker of alcohol self-association, which plays a crucial role in determining the thermodynamic properties of these solutions.

### 5.5. Results from the Flory model

Firstly, the low values of $\sigma_{\mathrm{r}}(H_{\mathrm{m}}^{\mathrm{E}})$ for mixtures with linear polyethers are noteworthy (Table 6). This indicates that the random mixing hypothesis is largely valid for these systems, which exhibit weak orientational effects. In other words, physical interactions dominate. In contrast, the $\sigma_{\mathrm{r}}(H_{\mathrm{m}}^{\mathrm{E}})$ result obtained for the (2-propanol + DPE) system is significantly higher, clearly indicating the importance of orientational effects in this solution. Notably, the $H_{\mathrm{m}}^{\mathrm{E}}$ curve of this system is skewed toward low mole fractions of the alkanol (Figure 4), a typical feature of solutions in which alcohol self-association plays a determining role.

It should also be noted that there is a significant difference between the values of the interactional parameter, $X_{12}$, for systems containing polyethers compared to the corresponding value for the solution with DPE (Table 6). This highlights that interactions between like molecules are much more prominent in the former mixtures.

The excess molar volumes are poorly represented by the model, with theoretical results significantly exceeding the experimental values. This indicates that the model overestimates the interactional contribution to the excess volume. Nevertheless, the theory correctly describes the trend observed in mixtures with polyethers: the decrease in $V_{\mathrm{m}}^{\mathrm{E}}$ as the number of O atoms in the ether increases.

Finally, it should be emphasized that the model provides similar results for (1-propanol or 2-propanol + glyme) mixtures. Accordingly, the mean relative standard deviations calculated from the formula:

$$\bar{\sigma}_r(H_m^E) = \frac{1}{N_s} \sum \sigma_r(H_m^E) \qquad (18)$$

are 0.040 (1-propanol) [8]; 0.044 (2-propanol) (with $N_s$ = 4, the number of systems considered). Consequently, orientational effects are also weak in systems containing 1-propanol.

## 6. Conclusions

Densities at atmospheric pressure and temperatures $T$ = (293.15 to 303.15) K have been measured for the systems [2-propanol + (CH$_3$O(CH$_2$CH$_2$O)$_u$CH$_3$] with $u$ = 1, 2, 3, 4. Excess molar enthalpies at $T$ = 298.15 K and the same pressure are also reported. The values of $H_m^E$ are large and positive, indicating that the disruption of interactions between molecules of the same species is the dominant contribution to this excess function. In contrast, the corresponding $V_m^E$ values are small or even negative (in the case of the solution with tetraglyme), suggesting that $V_m^E$ is primarily determined by structural effects. Analysis of the results using the Flory model reveals that orientational effects are weak in the studied systems and significantly stronger in the (2-propanol + di-$n$-propyl ether) mixture. Mixtures of 1-propanol or 2-propanol and glymes exhibit similar behavior, although interactions between unlike molecules are stronger in the solutions containing 1-propanol.

## Acknowledgements

J. V. A.-L. would like to thank the Instituto de Corresponsabilidade pela Educação (ICE) – Brazil for his PhD scholarship. F. P. acknowledges the FPI grant PREP2022-000047 from MCIN/AEI/10.13039/501100011033/ and FEDER, UE.

## Funding

This work was supported by Project PID2022-137104NA-I00 funded by MICIU/AEI/10.13039/501100011033 and by FEDER, UE.

# Density and excess molar enthalpy of (2-propanol + glyme) liquid mixtures. Application of the Flory model

## Supplementary Material

João Victor Alves-Laurentino[a], Fatemeh Pazoki[a], Luis Felipe Sanz[a], Juan Antonio González[a], Fernando Hevia[a], Daniel Lozano-Martín*,[a]

[a] GETEF. Departamento de Física Aplicada. Facultad de Ciencias. Universidad de Valladolid. Paseo de Belén, 7, 47011 Valladolid, Spain.

* Corresponding author, e-mail: daniel.lozano@uva.es

In this supplementary material, we report the excess partial molar volumes and enthalpies of both components of the (2-propanol + glyme) liquid mixtures studied (Tables S1 and S2, Figures S1-S4) at $T$ = 298.15 K. They can be obtained from the Redlich-Kister coefficients reported in Table 5 of the present paper using the following expressions:

$$\bar{F}_{m,1}^{E} = x_2^2 \sum_{i=0}^{k-1} A_i (2x_1 - 1)^i + 2x_2^2 x_1 \sum_{i=1}^{k-1} i A_i (2x_1 - 1)^{i-1}$$

$$\bar{F}_{m,2}^{E} = x_1^2 \sum_{i=0}^{k-1} A_i (2x_1 - 1)^i - 2x_1^2 x_2 \sum_{i=1}^{k-1} i A_i (2x_1 - 1)^{i-1}$$

where $F$ represents volume or enthalpy. We also report the partial molar volumes at $T$ = 298.15 K in Table S1, calculated from the expressions:

$$\bar{V}_{m,1} = V_{m,1} + \bar{V}_{m,1}^{E}$$

$$\bar{V}_{m,2} = V_{m,2} + \bar{V}_{m,2}^{E}$$

**Table S1.** Partial molar volumes, $\bar{V}_{m,1}$ and $\bar{V}_{m,2}$, and excess partial molar volumes, $\bar{V}_{m,1}^E$ and $\bar{V}_{m,2}^E$, at the temperature $T$ = 298.15 K, and pressure $p$ = 0.1 MPa, as function of the mole fraction of 2-propanol, $x_1$, for the (2-propanol (1) + linear polyether (2)) liquid mixtures.

| $x_1$ | $\bar{V}_{m,1}$ / (cm³ mol⁻¹) | $\bar{V}_{m,1}^E$ / (cm³ mol⁻¹) | $\bar{V}_{m,2}$ / (cm³ mol⁻¹) | $\bar{V}_{m,2}^E$ / (cm³ mol⁻¹) |
|---|---|---|---|---|
| | | 2-propanol (1) + monoglyme (2) | | |
| 0.05178 | 77.566 | 0.5982 | 104.585 | 0.0015 |
| 0.10439 | 77.510 | 0.5421 | 104.590 | 0.0063 |
| 0.15427 | 77.458 | 0.4906 | 104.598 | 0.0139 |
| 0.20588 | 77.406 | 0.4390 | 104.609 | 0.0253 |
| 0.30275 | 77.315 | 0.3479 | 104.640 | 0.0564 |
| 0.40779 | 77.226 | 0.2583 | 104.690 | 0.1059 |
| 0.49539 | 77.159 | 0.1920 | 104.744 | 0.1605 |
| 0.60142 | 77.091 | 0.1232 | 104.828 | 0.2443 |
| 0.70143 | 77.038 | 0.0709 | 104.926 | 0.3421 |
| 0.79875 | 77.000 | 0.0330 | 105.040 | 0.4560 |
| 0.85010 | 76.986 | 0.0185 | 105.108 | 0.5240 |
| 0.89810 | 76.976 | 0.0087 | 105.176 | 0.5925 |
| 0.95061 | 76.969 | 0.0021 | 105.257 | 0.6733 |
| | | 2-propanol (1) + diglyme (2) | | |
| 0.05109 | 77.203 | 0.2352 | 142.935 | −0.0011 |
| 0.10475 | 77.229 | 0.2618 | 142.933 | −0.0032 |
| 0.15607 | 77.238 | 0.2707 | 142.932 | −0.0045 |
| 0.19889 | 77.236 | 0.2682 | 142.932 | −0.0039 |
| 0.29787 | 77.205 | 0.2375 | 142.943 | 0.0067 |
| 0.39453 | 77.155 | 0.1878 | 142.970 | 0.0334 |
| 0.49704 | 77.097 | 0.1292 | 143.017 | 0.0808 |
| 0.59926 | 77.043 | 0.0761 | 143.082 | 0.1454 |
| 0.70252 | 77.003 | 0.0355 | 143.157 | 0.2208 |
| 0.79283 | 76.981 | 0.0132 | 143.222 | 0.2862 |
| 0.84771 | 76.973 | 0.0055 | 143.257 | 0.3211 |
| 0.89995 | 76.969 | 0.0016 | 143.284 | 0.3476 |
| 0.94705 | 76.968 | 0.0002 | 143.300 | 0.3637 |
| | | 2-propanol (1) + triglyme (2) | | |
| 0.10260 | 77.052 | 0.0845 | 181.850 | 0.0016 |
| 0.16302 | 77.037 | 0.0698 | 181.853 | 0.0038 |
| 0.21067 | 77.027 | 0.0595 | 181.855 | 0.0062 |
| 0.30289 | 77.010 | 0.0424 | 181.861 | 0.0120 |
| 0.40500 | 76.995 | 0.0277 | 181.869 | 0.0201 |
| 0.49568 | 76.985 | 0.0179 | 181.877 | 0.0281 |
| 0.59697 | 76.977 | 0.0100 | 181.886 | 0.0376 |
| 0.68977 | 76.973 | 0.0051 | 181.895 | 0.0463 |
| 0.79723 | 76.969 | 0.0018 | 181.905 | 0.0558 |
| 0.84925 | 76.968 | 0.0009 | 181.909 | 0.0601 |
| 0.89995 | 76.968 | 0.0003 | 181.913 | 0.0638 |
| | | 2-propanol (1) + tetraglyme (2) | | |
| 0.05776 | 76.850 | −0.1172 | 220.896 | −0.0016 |
| 0.10331 | 76.878 | −0.0895 | 220.893 | −0.0039 |
| 0.16636 | 76.897 | −0.0700 | 220.890 | −0.0069 |
| 0.19980 | 76.901 | −0.0663 | 220.890 | −0.0077 |
| 0.29568 | 76.896 | −0.0717 | 220.892 | −0.0056 |
| 0.40048 | 76.879 | −0.0881 | 220.900 | 0.0032 |
| 0.49985 | 76.869 | −0.0985 | 220.909 | 0.0115 |
| 0.59618 | 76.872 | −0.0958 | 220.905 | 0.0077 |
| 0.69890 | 76.890 | −0.0769 | 220.869 | −0.0285 |
| 0.79196 | 76.918 | −0.0490 | 220.785 | −0.1123 |
| 0.84810 | 76.937 | −0.0306 | 220.700 | −0.1970 |
| 0.90240 | 76.953 | −0.0146 | 220.587 | −0.3107 |
| 0.94957 | 76.963 | −0.0044 | 220.458 | −0.4396 |

**Table S2.** Excess partial molar enthalpies, $\bar{H}^E_{m,1}$ and $\bar{H}^E_{m,2}$, at the temperature $T$ = 298.15 K, and pressure $p$ = 0.1 MPa, as function of the mole fraction of 2-propanol, $x_1$, for the (2-propanol (1) + linear polyether (2)) liquid mixtures.

| $x_1$ | $\bar{H}^E_{m,1}$ / (J mol$^{-1}$) | $\bar{H}^E_{m,2}$ / (J mol$^{-1}$) |
|---|---|---|
| 2-propanol (1) + monoglyme (2) | | |
| 0.05116 | 4925 | 18 |
| 0.10266 | 4316 | 68 |
| 0.14753 | 3841 | 136 |
| 0.20743 | 3277 | 257 |
| 0.31511 | 2431 | 556 |
| 0.39259 | 1929 | 830 |
| 0.50397 | 1328 | 1318 |
| 0.60170 | 896 | 1853 |
| 0.71260 | 499 | 2616 |
| 0.79651 | 266 | 3335 |
| 0.85240 | 146 | 3898 |
| 0.89572 | 76 | 4389 |
| 0.93882 | 27 | 4930 |
| 2-propanol (1) + diglyme (2) | | |
| 0.05862 | 5273 | 20 |
| 0.09944 | 4858 | 55 |
| 0.14873 | 4399 | 120 |
| 0.19941 | 3967 | 211 |
| 0.29935 | 3207 | 464 |
| 0.39978 | 2531 | 828 |
| 0.50348 | 1895 | 1353 |
| 0.59184 | 1395 | 1961 |
| 0.69626 | 861 | 2933 |
| 0.80243 | 408 | 4296 |
| 0.85192 | 242 | 5094 |
| 0.89319 | 132 | 5851 |
| 0.97297 | 9 | 7586 |
| 2-propanol (1) + triglyme (2) | | |
| 0.04112 | 5498 | 1 |
| 0.10454 | 5245 | 22 |
| 0.15061 | 4953 | 65 |
| 0.19644 | 4611 | 138 |
| 0.29930 | 3769 | 417 |
| 0.40307 | 2948 | 862 |
| 0.50693 | 2223 | 1469 |
| 0.59909 | 1658 | 2170 |
| 0.69581 | 1121 | 3162 |
| 0.80344 | 580 | 4806 |
| 0.85000 | 374 | 5794 |
| 0.89311 | 209 | 6918 |
| 0.95217 | 48 | 8870 |
| 2-propanol (1) + tetraglyme (2) | | |
| 0.04079 | 5743 | −2 |
| 0.10101 | 5610 | 10 |
| 0.14880 | 5332 | 51 |
| 0.20190 | 4922 | 138 |
| 0.30358 | 4038 | 440 |
| 0.40874 | 3184 | 912 |
| 0.49262 | 2601 | 1391 |
| 0.59730 | 1967 | 2153 |
| 0.70215 | 1364 | 3284 |
| 0.80673 | 755 | 5194 |
| 0.84681 | 530 | 6274 |
| 0.90341 | 247 | 8280 |
| 0.95277 | 68 | 10621 |

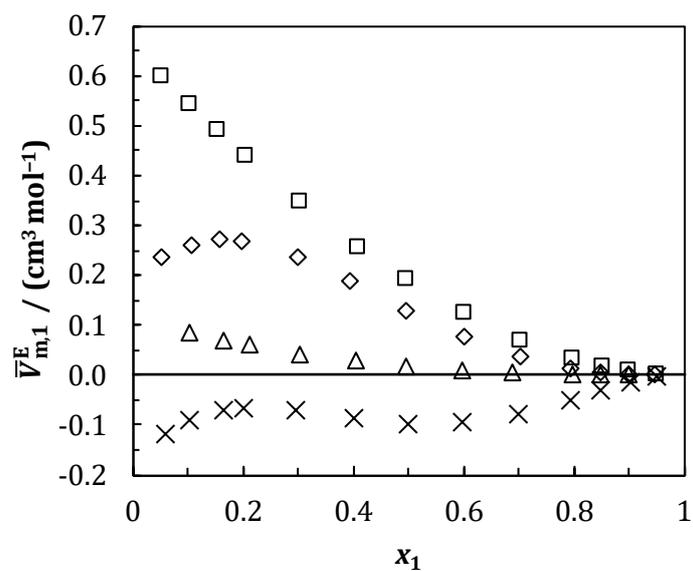

**Figure S1.** Excess partial molar volume, $\bar{V}_{m,1}^{E}$, of 2-propanol (1) as a function of the 2-propanol mole fraction, $x_1$, at temperature $T$ = 298.15 K and pressure $p$ = 0.1 MPa. Symbols: (□) (2-propanol + monoglyme), (◇) (2-propanol + diglyme), (△) (2-propanol + triglyme), (×) (2-propanol + tetraglyme).

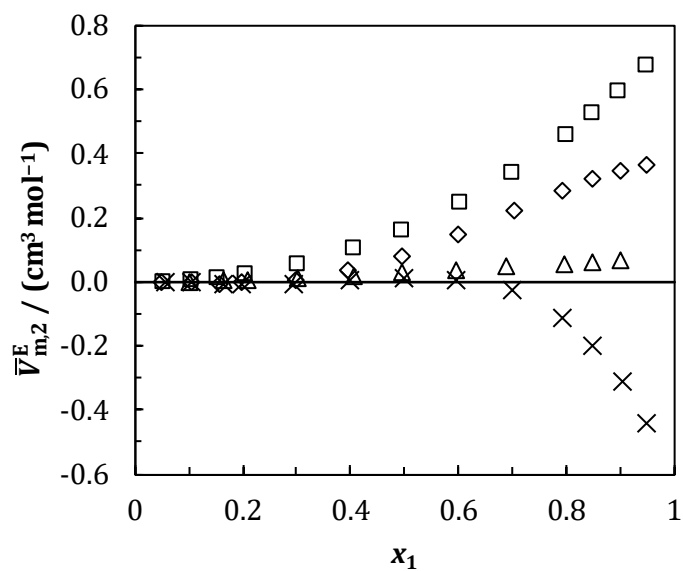

**Figure S2.** Excess partial molar volume, $\bar{V}_{m,2}^{E}$, of linear polyether (2) as a function of the 2-propanol mole fraction, $x_1$, at temperature $T$ = 298.15 K and pressure $p$ = 0.1 MPa. Symbols: (□) (2-propanol + monoglyme), (◇) (2-propanol + diglyme), (△) (2-propanol + triglyme), (×) (2-propanol + tetraglyme).

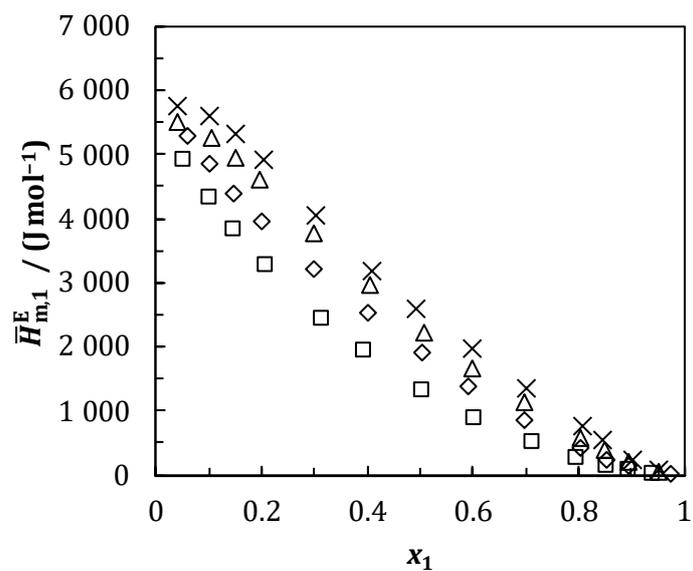

**Figure S3.** Excess partial molar enthalpy, $\bar{H}_{m,1}^{E}$, of 2-propanol (1) as a function of the 2-propanol mole fraction, $x_1$, at temperature $T$ = 298.15 K and pressure $p$ = 0.1 MPa. Symbols: (□) (2-propanol + monoglyme), (◇) (2-propanol + diglyme), (△) (2-propanol + triglyme), (×) (2-propanol + tetraglyme).

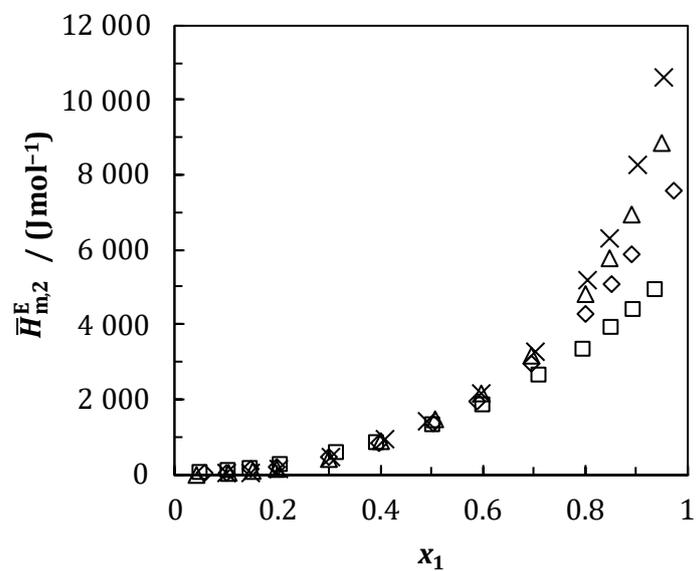

**Figure S4.** Excess partial molar volume, $\bar{H}_{m,2}^{E}$, of linear polyether (2) as a function of the 2-propanol mole fraction, $x_1$, at temperature $T$ = 298.15 K and pressure $p$ = 0.1 MPa. Symbols: (□) (2-propanol + monoglyme), (◇) (2-propanol + diglyme), (△) (2-propanol + triglyme), (×) (2-propanol + tetraglyme).

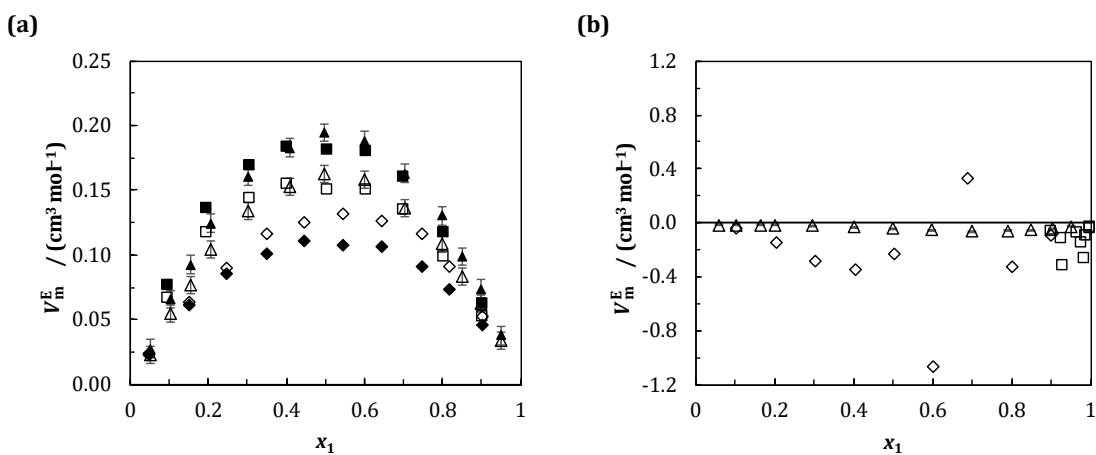

**Figure S5. (a)** Excess molar volume, $V_m^E$, as a function of the 2-propanol mole fraction, $x_1$, at temperatures $T$ = 293.15 K and 303.15 K for (2-propanol + monoglyme) mixture and pressure $p$ = 0.1 MPa. Symbols: (□) Benkelfat-Seladji et al. at 293.15 K [48], (◇) Almasi et al. at 293.15 K [51], (△) this work at 293.15 K, (■) Benkelfat-Seladji et al. at 303.15 K [48], (◆) Almasi et al. at 303.15 K [51], (▲) this work at 303.15 K. **(b)** Excess molar volume, $V_m^E$, as a function of the 2-propanol mole fraction, $x_1$, at temperature $T$ = 298.15 K for (2-propanol + tetraglyme) mixture and pressure $p$ = 0.1 MPa. Symbols: (□) Aznarez et al. (2004) [49], (◇) Aznarez et al. (2006) [50], (△) this work. Error bars show the experimental uncertainty from this work, $U(V_m^E)$.